# Photonic quadrupole topological phases in zero-dimensional cavity with synthetic dimensions

Weixuan Zhang[1, 2] and Xiangdong Zhang[1, 2*]

[1]*Key Laboratory of advanced optoelectronic quantum architecture and measurements of Ministry of Education, School of Physics, Beijing Institute of Technology, 100081, Beijing, China;*
[2]*Beijing Key Laboratory of Nanophotonics & Ultrafine Optoelectronic Systems, Micro-nano Center, School of Physics, Beijing Institute of Technology, 100081, Beijing, China;*
*Corresponding author, E-Mail address: zhangxd@bit.edu.cn*

**Quadrupole topological insulator, which supports robust corner states, has been recently demonstrated in two-dimensional (2D) spatial lattices. Here, we design the first photonic quadrupole topological insulator in fully synthetic spaces with the utilization of 0D optical cavity. The frequency and orbital angular momentum (OAM) of light are used to form the 2D synthetic spaces. Four degenerate polarization states are mapped to the internal lattice sites within the unit cell. By suitably engineering the coupling between cavity modes with different frequencies, OAMs and polarizations, the ideal synthetic quadrupole topological insulator is obtained. By using the robust synthetic corner state, we present the possibility for achieving topological protection of multi-photon entangled states. Our designed synthetic photonic quadrupole insulator proposes a unique platform to investigate higher-order topological phases in lower-dimensional system and possesses potential applications in quantum information and optical communication.**

The exploration of topological physics in both solid materials[1,2] and classical waves systems[3-6] has become one of the most fascinating frontiers in recent years. Based on the bulk-boundary correspondence principle, most of the *D*-dimensional topological systems studied so far are always featured by the (*D*-1)-dimensional boundary state[7-22]. Recently, a new class of symmetry-protected higher-order topological insulators that possess lower-dimensional boundary states and obey a generalization of the standard bulk-boundary correspondence have been introduced[23-35]. One example is given by the quadrupole topological insulator, which can exhibit ($D$-2)-dimensional topological states protected by the quantized bulk quadrupole polarization. Motivated by this novel property, many experimental implementations of the 2$D$ quadrupole topological phase with 0$D$ corner states have been realized in microwave[36], phononic[37], photonic[38] and electrical circuit systems[39].

On the other hand, except for the real physical dimensions, various degrees of freedom in photonic systems, such as frequency[40-43], OAM[44, 45] and external structural parameters[46-49], are also able to be used to construct the synthetic space where the innovative light control and quantum information processing may been realized. Based on the synthetic dimension in photonics, we can easily investigate higher-dimensional physics with lower-dimensional structures. For example, the Bloch oscillation in synthetic space along the frequency axis has been realized using a single dynamically modulated ring resonator[50]. The effective magnetic field for photons in the 2$D$ synthetic space is achieved in one optical cavity, where the topologically protected one-way edge state along the OAM- or frequency-axis appears[51]. In addition, the 3$D$ topological phases, such as Weyl point and topological insulators, can be constructed using 2$D$ systems with an auxiliary synthetic dimension[52-54]. Furthermore, some novel topological phenomena beyond 3$D$, such as 4$D$ quantum Hall effects, can also be implemented with the help of synthetic dimensions[48]. In all existing works, only first-order topological phases have been investigated in the synthetic space. The scheme for marrying higher-order topological physics with synthetic spaces, which can find wide applications in robust manipulation of various degrees of freedom in photonics, has never been proposed.

In this work, we design a photonic quadrupole topological insulator in synthetic 2$D$ spaces with the utilization of dynamically modulated 0$D$ optical cavity. The frequency and OAM of light are used to form the 2$D$ synthetic spaces. Four polarization states are mapped to internal lattice sites within the unit cell. By using this synthetic corner state, we prove that the topological protection of multi-photon entangled states may be achieved. Our work presents a new way to investigate higher-order topological phases in lower-dimensional system and possesses potential applications in quantum information and optical communications.

## Results

**Design of photonic quadrupole topological insulators with synthetic dimensions.** To design the synthetic quadrupole topological insulators, it is important to map different degrees of freedom in photonic systems to the real-space tight-binding lattice model with quantized bulk quadrupole polarization. The general conceptual drawing is graphically shown in Fig. 1a. Here, the frequency and OAM of light are set as two-axes in synthetic spaces. Four linked polarization modes with the same frequency and OAM are mapped to the intra-cell coupling condition (enclosed by the red dotted box). Four polarization angles used here are set as θ=0deg (*H*-polarized state, purple site), 45deg (*D*-polarized state, orange site), 90deg (*V*-polarized state, red site) and 135deg (*D⁻*-polarized state, green site), respectively. Additionally, the connection between lattice sites with different frequencies (or OAMs) represents the inter-cell couplings (enclosed by the black dotted box). By suitably designing the coupling strength between optical states with different frequencies, OAMs and polarizations, the ideal quadrupole topological insulators in the synthetic space can be obtained.

In the following, we prove that the above proposed scheme can be realized with the utilization of 0*D* optical ring cavity in free spaces. Here, the designed optical cavity should support a set of equally spaced resonant modes ($\omega_n=\omega_0+n\Omega$) when the dispersion effect is ignored. $\omega_0$ is the frequency of the 0*th* mode. $\Omega=2\pi c/L \ll \omega_0$ is the free-spectral range of the optical cavity with the loop length being *L*. *n* is an integer marking the mode index in the frequency axis. In addition, all optical elements in the cavity are chosen to have cylindrical symmetry, where the degenerated Laguerre-Gaussian modes with different radial (*p*) and azimuthal (*l*) indexes exist[51]. Moreover, each cavity mode with fixed OAM and frequency should possess four polarization states (θ=0, 45, 90 and 135deg). Hence, the electric field *E* in this optical cavity can be expanded as:

$$E = \sum_{l,n} \sum_{\theta}^{0,45,90,135} c_{l,n,\theta} A_{l,n,\theta}(r) e^{-il\phi} e^{i(\frac{\omega_n}{c} z - \omega_n t)}, \tag{1}$$

where $C_{l,n,\theta}$ is the amplitude for the cavity mode at frequency ($\omega_n$), OAM (*l*) and polarization ($\theta$). $A_{l,n,\theta}(r)$ is the corresponding modal profile perpendicular to the direction of beam propagation. $\phi$ denotes the azimuthal coordinate. These different cavity modes ($C_{l,n,\theta}$) can be mapped to the synthetic lattice sites in Fig. 1a. In this case, the synthetic quadrupole topological insulator is able to be created by suitably engineering the coupling strength between these cavity modes. To fulfill the ideal coupling configuration, the large optical cavity consisting of three parts (polarization cavity, OAM cavity and frequency cavity) is designed, as shown in Fig. 1b.

The intra-cell coupling within the synthetic lattice model can be achieved by using the designed polarization cavity (enclosed by the red dot box), where the required polarization conversions at the fixed frequency and OAM can be realized. In the synthetic unit cell, the coupling is allowed between the sites from the set {H($\theta$=0deg), V($\theta$=90deg)} and the set {D($\theta$=45deg), D⁻($\theta$=135deg)} but not between the sites in the same set. In this case, only eight kinds of polarization coupling (H$\leftrightarrows$D, D$\leftrightarrows$V, V$\leftrightarrows$D⁻ and D⁻$\leftrightarrows$H) are allowed. Coupling here means polarization interconversion. To fulfill this requirement, eight sub-loops with the same length *L* (coupling phase being $e^{i\omega_0 L/c}$) are designed. Each loop contains a polarization selector (only specific polarization state can pass through marked by colored triangles) and a polarization rotator ($\theta \rightarrow \theta \pm 45$deg shown by colored squares). In this case, when the *H*-polarized cavity mode enters in the polarization cavity, it can only go through the purple polarization selector. Then, the *H*-polarized mode is converted to D (and D⁻)-polarized modes based on the polarization rotator, where the polarization angle is increased (decreased) by 45deg. After these two steps, the modal coupling of H→D and H→D⁻ can be realized. Similarly, other sub-loops with different polarization selection and rotation operations are used to realize polarization interconversion with entered polarization states being *D*, V and D⁻, respectively. It is important to note that the negative intra-cell coupling should be introduced when the polarization interconversion of "$D^- \underset{\rightarrow}{\leftarrow} H$" is carried out. This can be easily achieved by using a phase retarder (marked by yellow rectangles in Fig. 1b) with a π phase delay. Moreover, the reflectivities of different beam splitters (BSs) used here should be suitably engineered to make the absolute value of the intra-cell coupling coefficients become identical. Here, $R(BS_1)=0.25$, $R(BS_1)=0.33$ and $R(BS_3)=0.5$.

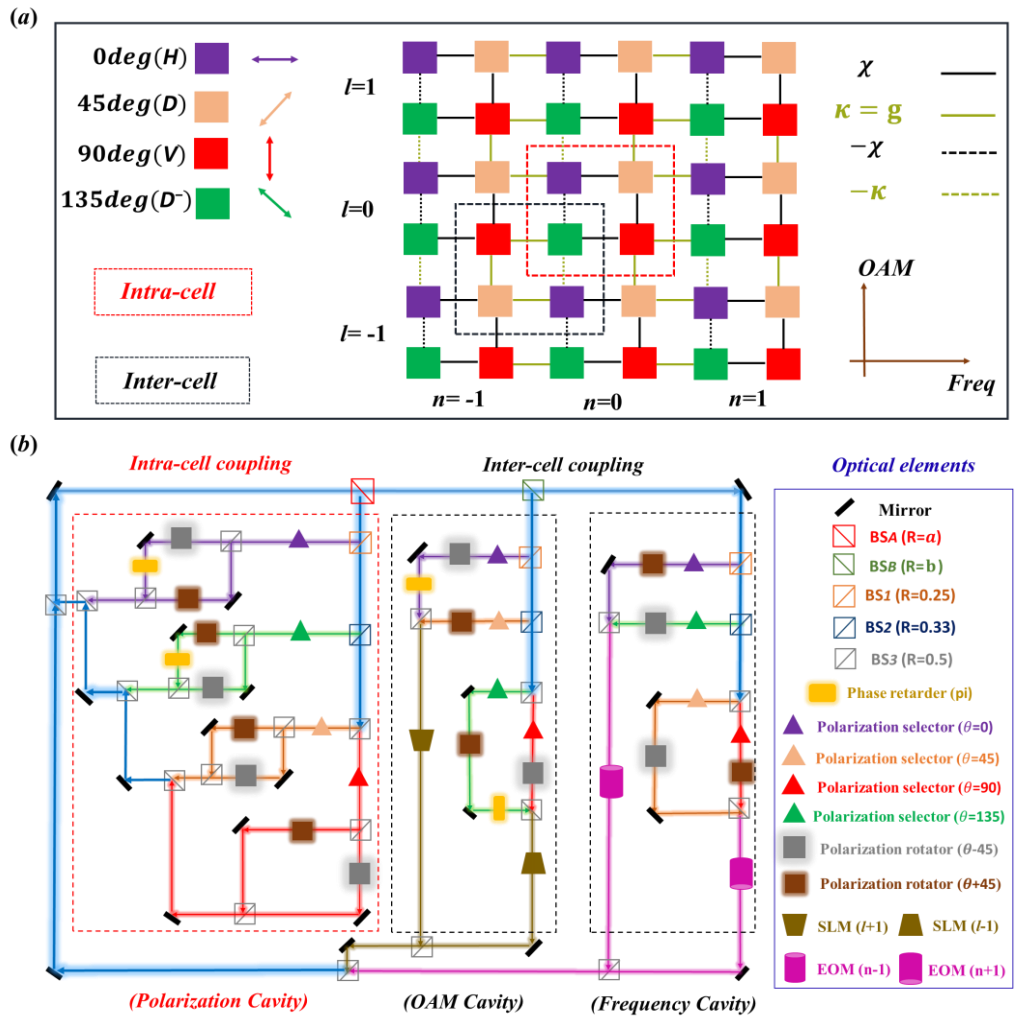

**Figure 1 | Design of synthetic quadrupole topological insulator in zero-dimensional cavity.** (a) The lattice model of quadrupole topological phases in synthetic 2D spaces. The frequency and OAM of light are set as two-axes in synthetic spaces. Four polarization states θ=0deg (*H*), 45deg (*D*), 90deg (*V*) and 135deg (*D*⁻) are considered as the internal lattice sites within the unit cell. The red and black dot boxes marked the intra- and inter-cell coupling conditions. (b) The designed 0D optical cavities, which include three cavities (polarization, OAM and frequency cavities), can realize the quadrupole topological insulator in synthetic spaces. EOM is the electro-optic phase modulator, SLM is the spatial light modulator, and BS is the beam splitter.

Based on the coupled mode theory, the change of the modal amplitude after each round-trip within the polarization cavity can be expressed as (the coupling phase

$e^{i\omega_0 L/c}$ is a global phase, which can be neglected without loss of generality):

$$\frac{dc_{l,n,0}}{dt}=i\chi(c_{l,n,45}+e^{i\pi}c_{l,n,135}), \quad \frac{dc_{l,n,45}}{dt}=i\chi(c_{l,n,90}+c_{l,n,0}),$$
$$\frac{dc_{l,n,90}}{dt}=i\chi(c_{l,n,135}+c_{l,n,45}), \quad \frac{dc_{l,n,135}}{dt}=i\chi(e^{i\pi}c_{l,n,0}+c_{l,n,90}), \quad (2)$$

where the value of coupling coefficient is expressed as $\chi=R(BS_A)*R(BS_1)*R(BS_3)=a/8$ ($a$ is the reflectivity of the 'BS$_A$'). It is clearly shown that our designed modal conversion configuration induced by the polarization cavity possesses an ideal correspondence to the intra-cell coupling of the lattice model with quantized bulk quadrupole polarization.

On the other hand, to introduce appropriate inter-cell couplings between different synthetic unit cells, another two cavities (enclosed by black dot boxes in Fig. 1b) with either OAM or frequency control should be designed. Firstly, we focus on the modal coupling with different OAMs and polarizations (but the same frequency) by using an OAM modulated cavity with four sub-loops. Here, only four kinds of modal couplings, that are $H\xrightarrow{l+1}D^-$, $D^-\xrightarrow{l-1}H$, $D\xrightarrow{l+1}V$ and $V\xrightarrow{l+1}D$, are allowed. The length of each loop is equal to $L$ so that the global coupling phase $e^{i\omega_0 L/c}$ can be neglected. To fulfill the coupling between cavity modes with different OAMs, each sub-loop should possess a spatial light modulator (SLM) to change the OAM of light by either +1 or -1. In addition, each sub-loop should also possess required polarization control to achieve the accompanied polarization conversions. For cases of $H\xrightarrow{l+1}D^-$ and $D^-\xrightarrow{l-1}H$, a π phase delay is applied to introduce the negative inter-cell couplings. In this condition, when the *V*-polarized fields enter in the OAM cavity, it can only pass through the red polarization selector. Then, the designed polarization rotator converts the *V*-polarized mode to the D-polarized mode. Finally, the OAM index of the D-polarized cavity mode decreases by one with the help of the SLM. Based on these three steps, the modal coupling of $V\xrightarrow{l-1}D$ can be realized. Similarly, other sub-loops are used to realize modal conversion with entered polarization states being *H*, *D* and *D*⁻, respectively.

Based on the above discussion, the variation of cavity modes with the light travelling a round-trip within the OAM cavity can be expressed as:

$$\frac{dc_{l,n,0}}{dt} = i\kappa e^{i\pi} c_{l+1,n,135}, \qquad \frac{dc_{l,n,45}}{dt} = i\kappa c_{l+1,n,90},$$
$$\frac{dc_{l,n,135}}{dt} = i\kappa e^{i\pi} c_{l-1,n,0}, \qquad \frac{dc_{l,n,90}}{dt} = i\kappa c_{l-1,n,45}, \qquad (3)$$

where the coupling coefficient can be expressed as $\kappa = [1-R(BS_A)]*R(BS_B)*R(BS_1) = (1-a)b/4$ ($b$ is the reflectivity of 'BS$_B$').

Except for the modal coupling with different OAMs, the conversion between cavity modes with different frequencies and polarizations (but the same OAM) should also be achieved. Here, four kinds of modal couplings, that are $H \xrightarrow{n-1} D$, $D^- \xrightarrow{n-1} V$, $D \xrightarrow{n+1} H$ and $V \xrightarrow{n+1} D^-$, are allowed. The length of each sub-loop in the frequency cavity is $L$. To realize modal coupling with different frequency, each sub-loop possesses an electro-optic phase modulators (EOMs) to induce time-dependent transmissions[50-53]. There are two different transmission coefficients induced by EOMs, which can be expressed as: $T = e^{i\beta[\cos(\Omega_M t) \pm i\sin(\Omega_M t)]}$, where $\Omega_M$ is the modulation frequency and $\beta$ denotes the modulation amplitude. In the presence of a travelling signal $e^{i\omega t}$, the transmitted wave can be expressed as: $Te^{i\omega t} \approx e^{i\omega t} + \beta e^{i(\omega \pm \Omega_M)t}$. It is clearly shown that the proposed EOM can change the modal frequency by either $\Omega_M$ or $-\Omega_M$. Combining this dynamically modal control and suitable polarization manipulations, the ideal inter-cell coupling along frequency dimension can be realized. For example, the $D^-$-polarized fields entering in the frequency cavity can only pass through the green polarization selector. Then, the designed polarization rotator can convert the $D^-$-polarized mode to the $V$-polarized mode. Finally, based on the dynamically modulated EOM, the frequency index of the V-polarized cavity mode changes to n-1. In this case, the modal coupling of $D^- \xrightarrow{n-1} V$ is fulfilled.

In the weak modulation regime and without the detuning of the modulation frequency from the free-spectral range of the cavity ($\Omega_M = \Omega$), the change of the modal

amplitude after each round-trip with light passing through the EOM gives:

$$\frac{dc_{l,n,0}}{dt}=igc_{l,n-1,45}, \qquad \frac{dc_{l,n,135}}{dt}=igc_{l,n-1,90},$$
$$\frac{dc_{l,n,45}}{dt}=igc_{l,n+1,0}, \qquad \frac{dc_{l,n,90}}{dt}=igc_{l,n+1,135} \tag{4}$$

Here, the coupling coefficient is expressed as $g=\beta[1-R(BS_A)]*[1-R(BS_B)]*R(BS_1)$ $=\beta(1-a)(1-b)/4$.

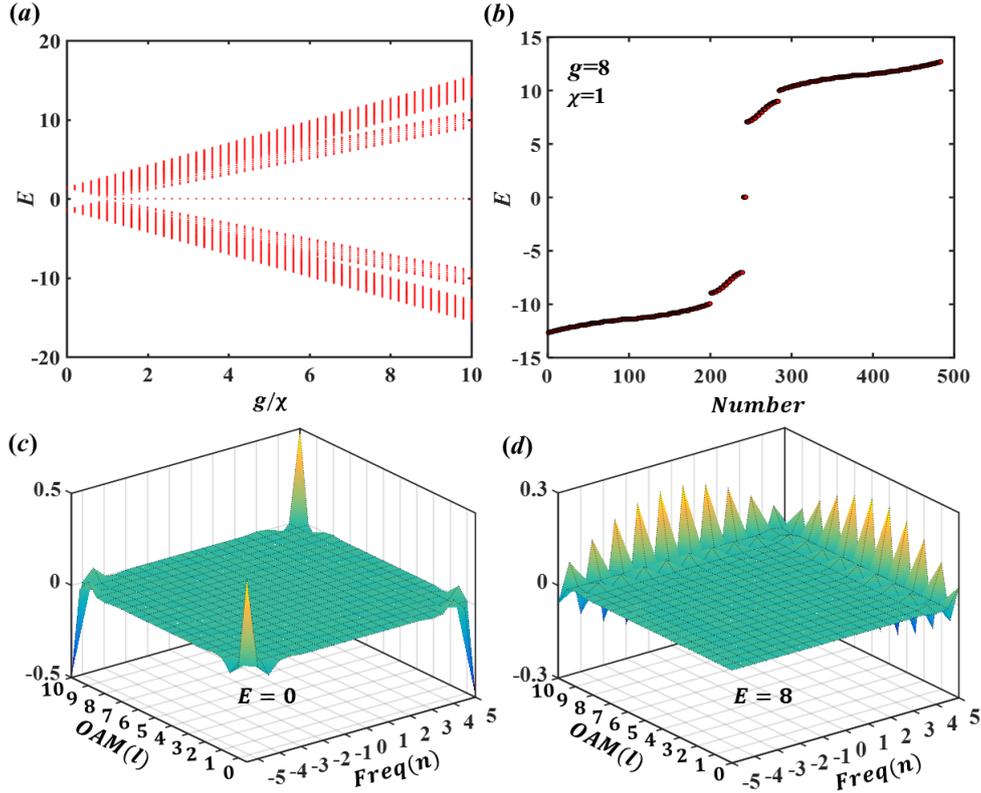

**Figure 2 | Eigenspectrum and eigenmodes for the synthetic quadrupole topological insulator.**
(a) The evolution of eigenspectrum with different values of $g/\chi$ ($\chi=1$). (b) The eigenspectrum at $g=8$. (c) and (d) The modal distribution of corner and edge states in the synthetic space.

Combing the modal coupling equations (*Eq*. 2-4), the Hamiltonian of such an optical cavity can be expressed as:

$$\begin{aligned}H=&\sum_{n,l}g(a^+_{l,n-1,45}a_{l,n,0}+a^+_{l,n-1,90}a_{l,n,135}+a^+_{l,n+1,0}a_{l,n,45}+a^+_{l,n+1,135}a_{l,n,90})\\&+\kappa\,(a^+_{l+1,n,135}a_{l,n,0}e^{i\pi}+a^+_{l+1,n,90}a_{l,n,45}+a^+_{l-1,n,0}a_{l,n,135}e^{i\pi}+a^+_{l-1,n,45}a_{l,n,90})\\&+\chi(a^+_{l,n,45}a_{l,n,0}+a^+_{l,n,90}a_{l,n,45}+a^+_{l,n,135}a_{l,n,90}+e^{i\pi}a^+_{l,n,0}a_{l,n,135}\\&+e^{i\pi}a^+_{l,n,135}a_{l,n,0}+a^+_{l,n,0}a_{l,n,45}+a^+_{l,n,45}a_{l,n,90}+a^+_{l,n,90}a_{l,n,135}),\end{aligned} \tag{5}$$

where we set $\hbar=1$. $a^+_{l,n,\theta}$ ($a_{l,n,\theta}$) is the creation (annihilation) operator for the cavity mode with frequency $n$, OAM $l$ and polarization $\theta$. The Hamiltonian in equation 4 presents the same form with that of the quadrupole topological insulator in real space[24,25]. In following parts, we assume that inter-cell couplings along frequency and OAM axes are identical $g=\kappa$. This can be realized when the modulation strength and reflectivity of 'BS$_B$' satisfy $b=\beta/(1+\beta)$.

In order to exhibit the higher-order topological effect in our designed synthetic space, the artificial boundaries along both frequency and OAM axes should be included. This is not a trivial problem since the boundaries are not real. We can follow the previous work to introduce the frequency boundary by coupling with a secondary cavity as proposed in ref.[51]. And, the key of creating the OAM boundary is the design of holes in BSs, which is systematically analyzed in ref.[45]. Actually, there are many factors limiting the size of the synthetic lattice, for example, the effective operation range of the optical elements. Here, we assume that frequency boundaries locate at $n=\pm 5$ and OAM boundaries are chosen as $l=0$ and $l=10$. In this case, the finite lattice model with open boundary condition is generated in the synthetic space where totally 484 sites (11 resonant frequencies, 11 OAMs and 4 polarizations) exist in this system. To illustrate the relationship between the topological property and the ratio between inter- and intra-cell coupling in synthetic dimensions, we plot the evolution of eigenspectrum of Hamiltonian (equation 5) with different values of $g/\chi$ ($\chi$ is normalized to be 1), as shown in Fig. 2a. For the practical realization of these different coupling relations, the reflectivities of "BS$_A$" and "BS$_B$" should be suitably adjusted. We find that there is neither edge nor corner state within the trivial bulk band gap when $g<\chi$ [$b<0.5a/(1-a)$]. And, the topological phase transition appears at the point with $g=\chi$ where the bulk band gap gets closed. Beyond the transition point ($g>\chi$), the bulk gap opens again. Then, both edge and midgap corner states appear. This can be further shown in Fig. 2b where the corresponding eigenvalues at $g=8$ [$b=4a/(1-a)$] are plotted. It is find that both fourfold degenerate 'zero-energy' corner states and gapped edge states exist in this topologically non-trivial band gap. To clearly illustrate the midgap corner and edge states, in Fig. 2c and 2d, the distribution of the corner ($E$=0) and edge ($E$=8) states in the synthetic space are plotted. As expected, the eigenmode is highly localized at corners or edges within the finite lattice model in synthetic spaces that is the same as the quadrupole topological insular in real spaces.

**Topologically protected evolution of corner state in synthetic spaces.**

To characterize the evolution of corner and edge states in the synthetic space, we perform numerical simulation of output fields from the designed optical cavity. The general coupled-mode equations for our designed synthetic quadrupole topological insulators with the consideration of input and output ports can be expressed as:

$$\frac{dc_{l,n,0}}{dt} = -\gamma c_{l,n,0} - i\chi c_{l,n,45} + i\chi c_{l,n,135} - igc_{l,n-1,45}(1-\delta_{-5,n}) - i\kappa c_{l+1,n,135}(1-\delta_{10,l}) + \gamma^{1/2} S^{in}_{l,n,0}$$

$$\frac{dc_{l,n,45}}{dt} = -\gamma c_{l,n,45} - i\chi c_{l,n,0} - i\chi c_{l,n,90} - igc_{l,n+1,0}(1-\delta_{5,n}) - i\kappa c_{l+1,n,90}(1-\delta_{10,l})$$

$$\frac{dc_{l,n,90}}{dt} = -\gamma c_{l,n,90} - i\chi c_{l,n,45} - i\chi c_{l,n,135} - igc_{l,n+1,135}(1-\delta_{5,n}) - i\kappa c_{l-1,n,45}(1-\delta_{0,l}) \qquad (6)$$

$$\frac{dc_{l,n,135}}{dt} = -\gamma c_{l,n,135} - i\chi c_{l,n,90} + i\chi c_{l,n,0} - igc_{l,n-1,90}(1-\delta_{-5,n}) - i\kappa c_{l-1,n,0}(1-\delta_{0,l})$$

$$S^{out}_{l,n,\theta} = \gamma^{1/2} c_{l,n,\theta} \ ,$$

where $\gamma/2$ is the decay rate through the coupling between the optical cavity and the input-output channel. We assume that all cavity modes couple to the input-output channel with equal rates. $S^{in}_{l,n,\theta}$ ($S^{out}_{l,n,\theta}$) is the input (output) modal amplitude at frequency $n$, OAM $l$ and polarization $\theta$. The artificial boundary in the synthetic space is chosen as $n=\pm 5$, $l=0$ and $l=10$. Here, we only focus on the case that the polarization angle of input signal is 0deg. Similar results with other input polarizations are provided in the Supplementary Information.

In our simulations, we set the coupling rate between the input (and output) port and the optical cavity as $\gamma$=0.1. And, all other parameters are identical with those used in the calculation of eigenspectrum shown in Fig. 2a. To effectively excite the synthetic corner state, the input signal should locate at the corner (left-up) of the synthetic space. No corner excitation appears when the input state locates in the bulk of the synthetic space (see Supplementary Information). Figs. 3a-3c show the steady-state modal distribution with the value of $g(\chi)$ being 1(1), 2(1) and 8(1), respectively. The input signal is selected as $S_{10,-5,0}^{in} = e^{i0t}$ ('zero energy' excitation). We find that no corner localization appears when the coupling strength satisfied the relationship of $g=\chi$. This is due to the fact that no band gap exists in this case. While, with increasing the value of $g/\chi$, the concentration of the output optical mode at the corner of the synthetic space appears. Additionally, the modal profile is completely the same as the midgap corner state, as shown in Fig. 2c. Moreover, the larger the value of $g/\chi$ is, the more significant localization appears at the corner of the synthetic space. If $g/\chi<1$, corner states cannot be excited in the trivial band gap (see Supplementary Information). Except for the corner excitation, the synthetic edge state can also be excited with suitable input signals (see Supplemental Information).

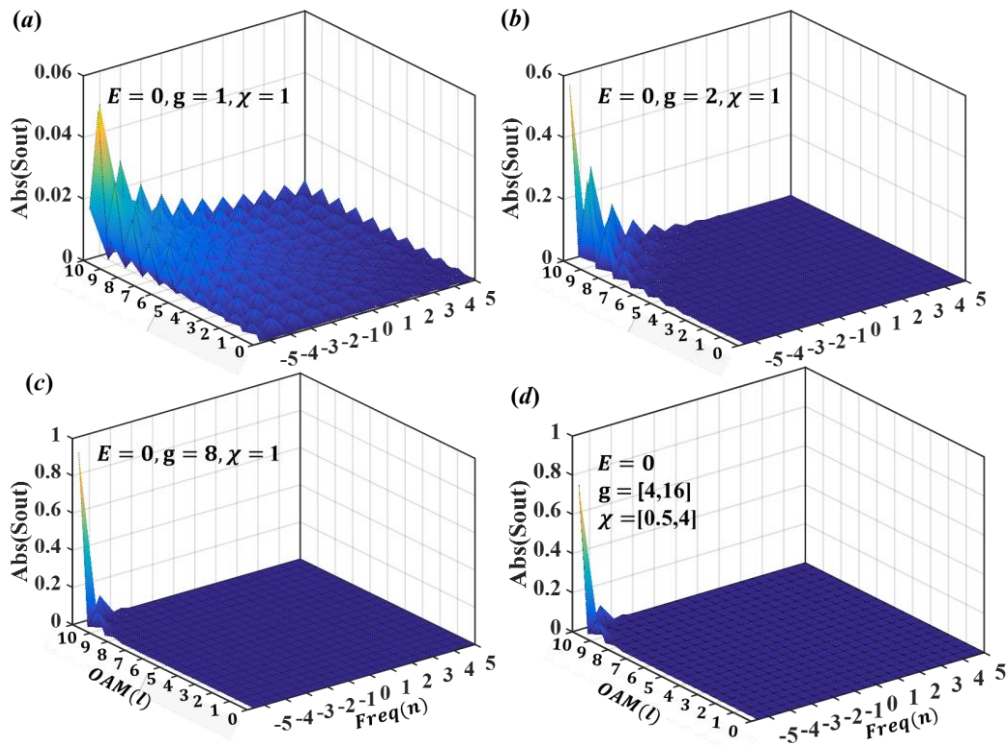

**Figure 3 | Evolution of corner state in synthetic spaces.** (a)-(c) The steady-state modal distribution with the value of $g(\chi)$ being 1(1), 2(1) and 8(1), respectively. (d) The steady-state modal distribution of corner state under the influence of disorder where $g(\chi)$ are randomly selected ranging from 4(0.5) to 16(4) between different lattice sites.

The most important property of the topologically-protected corner state is that it is robust with disorders. To illustrate this phenomena, we calculate the steady-state solution of the output field with significant disorders on $g/\chi$ at different lattice sites. The configuration average (50 times) is performed to eliminate the accidental result in the disorder system. The results are shown in Fig. 3d. We find that sharply localized output fields at the corner of the synthetic space still exist at zero energy when the values of $g(\chi)$ on different lattice sites are randomly selected ranging from 4(0.5) to 16(4). It clearly reveals the robustness of the corner state for the synthetic quadrupole topological insulator.

**Topological protection of multi-photon entangled states with synthetic corner states.** A fascinated potential application of our designed synthetic quadrupole topological insulator is the realization of topological protection of multi-photon entangled states. Creation and manipulation of entangled optical states have been extensively investigated in recent years. However, many errors arising from random disorders limit their efficiencies. The discoveries regarding topological phases have introduced avenues to construct quantum optical systems that are protected against imperfections. For example, the topological propagation of bi-photon state has been realized in a nanophotonics lattice[55]. In addition, using traditional optical elements to control quantum states is very important in modern optical communications and quantum information. Consequently, the robust propagation of multi-photon entangled states by conventional optical devices may possess potential applications in future all-optical quantum network.

In the following, we show that the topologically protected entanglement between two, three and four photons may be realized based on the designed synthetic corner state. To illustrate this effect, we focus on the quantum state composed of four photons, where the frequencies, OAMs and polarizations are both entangled. The results for two and three photons are provided in the Supplementary Information. The single-photon state should be selected at four corners in the synthetic space. These four corner states can be expressed as: $\varphi_1(H, l=10, n=-5)$, $\varphi_2(D, l=10, n=5)$, $\varphi_3(V, l=0, n=5)$ and $\varphi_4(D^-, l=0, n=-5)$. Here, as an example, the input 4-photon entangled state can be written as:

$$|\psi\rangle = \frac{1}{\sqrt{2}}(|H,L=10,n=-5\rangle|D,L=10,n=5\rangle|V,L=0,n=5\rangle|D^-,L=0,n=-5\rangle \quad (7)$$
$$+|V,L=0,n=5\rangle|D^-,L=0,n=-5\rangle|H,L=10,n=5\rangle|D,L=10,n=5\rangle)$$

In Fig. 4a, we plot the numerical result of the output 4-photon steady-state with "zero energy" excitation when the systematic parameters are the same to those used in Fig. 3c. It is clearly shown that output photons are still located at the corner in the synthetic space so that both the polarization, OAM and frequency entanglements are preserved with a high fidelity (~92%, peak value of the output field at the corner in the synthetic space). We have also designed the scheme of quantum state tomography for the entangled state (see Supplementary Information for details).

Moreover, when significant disorders are introduced in this synthetic quadrupole topological phase [$g(\chi)$ are randomly selected ranging from 4(0.5) to 16(4) on different lattice sites], as shown in Fig. 4b, we find that the output modes of four photons are still mainly concentrated at four corners in the synthetic space (the fidelity is nearly 75%). The configuration average (50 times) has been performed. Consequently, we note that the entangled multi-photon state is robust with disorders within the synthetic quadrupole topological insulators.

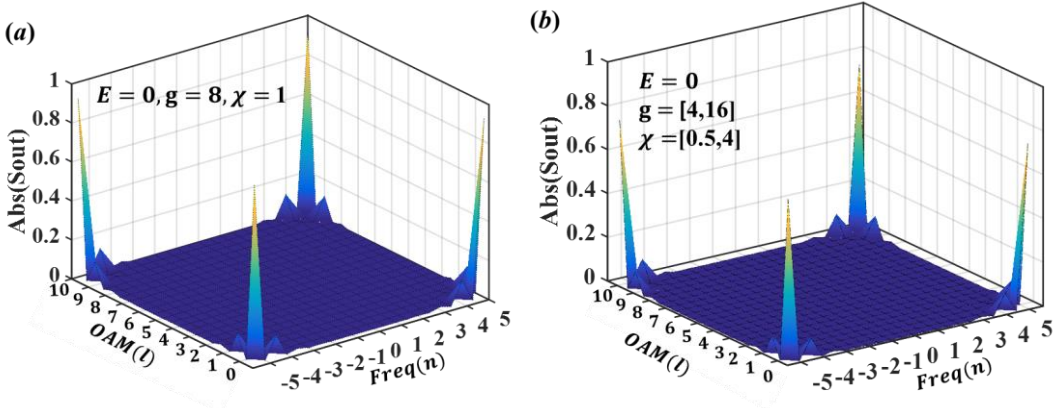

**Figure 4 | Evolution of multi-photon entangled states with the excitation of synthetic corner state.** (a) The steady-state solution of the output four photons state with the value of $g(\chi)$ being 8(1). (b) The steady-state modal distribution of output four photons state under the influence of disorder where $g(\chi)$ are randomly selected ranging from 4(0.5) to 16(4) on different lattice sites.

**Discussion.** Although the theoretical aspects of synthetic quadrupole topological insulators are only discussed in the present work, the designed 0D optical cavity with frequency, OAM and polarization controls may be actually realized in experiments. In the free-space optics, we may let the central frequency $\omega_0$ and the length of the optical cavity $L$ be 100THz and 1~10m, respectively. The modal spacing and modulation frequency of EOM are both nearly ranging from 10 to 100MHz$\ll\omega_0$[51]. With such dense modal distribution, the dispersion effect can be neglected for a large number of frequencies of cavity modes. However, it is important to note that losses in all optical elements can lead to a broadening of the cavity mode, which can make the dense cavity modes become undistinguishable. To solve this problem, we can place an amplifier in the cavity loops to increase the quality factor (without lasing) of the cavity mode (decrease the linewidth of the cavity mode). It is well known that amplifiers can add noises. Hence, to weak the noise, we can increase the gain factor and adjust the impedance of the source. Additionally, due to the topological robustness of this synthetic quadrupole insulator, the entanglement state can still possess high quality even slightly noises exist. On the other hand, to realize the non-trivial coupling conditions ($g=\kappa>\chi$), the reflectivities of 'BS$_A$' and 'BS$_B$' should be suitably tuned to satisfy the relationship of $b=\beta/(1+\beta)>0.5a/(1-a)$. In the weak modulation range $\beta\ll1$, the reflectivities of 'BS$_A$' and 'BS$_B$' are also very low. The most difficult procedure for the realization of corner state is introducing "boundaries" in the synthetic space. We can follow the previous work to introduce the frequency boundary by coupling with a secondary cavity[51], and to create the OAM boundary by designing holes in BSs[45].

    In conclusion, we have theoretically designed the first photonic quadrupole topological insulator in 2$D$ synthetic spaces with the utilization of 0$D$ optical cavities.

The coupling between optical cavity modes with different frequencies, OAMs and polarizations are used to fulfill the corresponding tight-binding lattice model in the synthetic space. The numerical results based on the coupled-mode theory demonstrated that the steady-state solution at "zero energy" of the output field from the synthetic quadrupole topological insulator is totally located at the corner of the 2*D* synthetic spaces even if significant disorders exist. By using the robust corner state in the synthetic spaces, the topological protection of multi-photon entangled state is proved. Our proposed method may be used to realize many other synthetic higher-order topological phases, such as octupole insulators, quadrupole semimetals and even 4D higher-order topological phases. Moreover, the proposed synthetic corner state may possess potential applications in quantum information processing and offer possibilities for robust shaping of light in synthetic dimensions.

**References**

1. Hasan, M. Z. & Kane, C. L. Colloquium: Topological insulators, *Rev. Mod. Phys.* **82**, 3045–3067 (2010).
2. Qi, X. L. & Zhang, S. C. Topological insulators and superconductors, *Rev. Mod. Phys.* **83**, 1057–1110 (2011).
3. Lu, L., Joannopoulos, J. D. & Soljačić, M. Topological photonics, *Nat. Photonics* **8**, 821-829 (2014).
4. Ozawa, T., Price, H. M., Amo, A., Goldman, N., Hafezi, M., Lu, L., Rechtsman, M. C., Schuster, D., Zilberberg, O. & Carusotto, L. Topological photonics, *Rev. Mod. Phys.* **91**, 015006 (2019).
5. Khanikaev, A.B. & Shvets, G. Two-dimensional topological photonics, *Nat. Photonics* **11**, 763-773 (2017).
6. Ma, G., Meng, X. & Chan, C. T. Topological phases in acoustic and mechanical systems, *Nat. Rev. Phys.* **1**, 281-294 (2019).
7. Su, W. P., Schrieffer, J. R. & Heeger, A. J. Solitons in polyacetylene, *Phys. Rev. Lett.* **42**, 1698-1701 (1979).
8. Meng, X., Zhang, Z. Q. & Chan, C. T. Surface impedance and bulk band geometric phases in one-dimensional systems, *Phys. Rev. X* **4**, 021017 (2014).

9. Blanco-Redondo, A., Andonegui, I., Collins, M. J., Harari, G., Lumer, Y., Rechtsman, M.C., Eggleton, B. J. & Segev, M. Topological optical waveguiding in silicon and the transition between topological and trivial defect states, *Phys. Rev. Lett.* **116**, 163901 (2016).

10. Slobozhanyuk, A. P., Poddubny, A. N., Miroshnichenko, A. E., Belov, P. A. & Kivshar, Y. S. Subwavelength topological edge states in optically resonant dielectric structures, *Phys. Rev. Lett.* **114**, 123901 (2015).

11. Zhang, W. & Zhang, X. Backscattering-Immune computing of spatial differentiation by nonreciprocal plasmonics, Phys. Rev. Applied **11**, 054003 (2019).

12. Poddubny, A., Miroshnichenko, A., Slobozhanyuk, A. & Kivshar, Y. S. Topological Majorana States in Zigzag Chains of Plasmonic Nanoparticles, *ACS Photonics* **3**, 1468 (2016).

13. Zhang, W., Wu, T. & Zhang, X. Reconfigurable topological phases in photoexcited graphene nanoribbon arrays, *J. Opt.* **20**, 095005 (2018).

14. Haldane, F. D. M. & Raghu, S. Possible realization of directional optical waveguides in photonic crystals with broken time-reversal symmetry, *Phys. Rev. Lett.* **100**, 013904 (2008).

15. Wang, Z., Chong, Y. D., Joannopoulos, J. D. & Soljačić, M. Observation of unidirectional backscattering-immune topological electromagnetic states, *Nature* **461**, 772-775 (2009).

16. Rechtsman, M. C., Zeuner, J. M., Plotnik, Y., Lumer, Y., Podolsky, D., Dreisow, F., Nolte, S., Segev, M. & Szameit, A. Photonic Floquet topological insulators, *Nature* **496**, 196-200 (2013).

17. Fang, K., Yu, Z. & Fan, S. Realizing effective magnetic field for photons by controlling the phase of dynamic modulation, *Nat. Photonics* **6**, 782-787 (2012).

18. Hafezi, M., Demler, E. A., Lukin, M. D. & Taylor, J. M. Robust optical delay lines with topological protection, *Nat. Phys.* **7**, 907-912 (2011).

19. Khanikaev, A. B., Mousavi, S. H., Tse, W. K., Kargarian, M., Macdonald, A. H. & Shvets, G. Photonic topological insulators, *Nat. Mater.* **12**, 233-239 (2013).

20. He, C., Sun, X. C., Liu, X. P., Lu, M. H., Chen, Y., Feng, L. & Chen, Y. F. Photonic topological insulators with broken time-reversal symmetry, *Proc. Natl.*

*Acad. Sci.* **113**, 4924-4928 (2016).

21. Wu, L. H. & Hu, X. Scheme for achieving a topological photonic crystal by using dielectric material, *Phys. Rev. Lett.* **114**, 223901 (2015).

22. Hafezi, M., Mittal, S., Fan, J., Migdall, A. & Taylor, J. M. Imaging topological edge states in silicon photonics, *Nat. Photonics* **7**, 1001-1005 (2013).

23. Zhang, F., Kane, C. L. & Mele, E. J. Surface State Magnetization and Chiral Edge States on Topological Insulators, *Phys. Rev. Lett.* **110**, 046404 (2013).

24. Benalcazar, W. A., Bernevig, B. A. & Hughes, T. L. Quantized electric multipole insulators, *Science* **357**, 61 (2017).

25. Benalcazar, W. A., Bernevig, B. A. & Hughes, T. L. Electric multipole moments, topological multipole moment pumping, and chiral hinge states in crystalline insulators, *Phys. Rev. B* **96**, 245115 (2017).

26. Langbehn, J., Peng, Y., Trifunovic, L., Oppen, F. & Brouwer, P. W. Reflection-Symmetric Second-Order Topological Insulators and Superconductors, *Phys. Rev. Lett.* **119**, 246401 (2017).

27. Schindler, F., Cook, A. M., Vergniory, M. G., Wang, Z., Parkin, S. S. P., Bernevig, B. A. & Neupert, T. Higher-order topological insulators, *Sci. Adv.* **4**, 0346 (2018).

28. Schindler, F., Wang, Z., Vergniory, M. G., Cook, A. M., Murani, A., Sengupta, S., Kasumov, A. Yu., Deblock, R., Jeon, S., Drozdov, I., Bouchiat, H., Guéron, S., Yazdani, A., Bernevig, B. A. & Neupert, T. Higher-order topology in bismuth, *Nat. Phys.* **14**, 918 (2018).

29. Song, Z., Fang, Z. & Fang, C. (d-2)-Dimensional Edge States of Rotation Symmetry Protected Topological States, *Phys. Rev. Lett.* **119**, 246402 (2017).

30. Ezawa, M. Higher-Order Topological Insulators and Semimetals on the Breathing Kagome and Pyrochlore Lattices, *Phys. Rev. Lett.* **120**, 026801 (2018).

31. Kunst, F. K., Miert, G. V. & Bergholtz, E. J. Lattice models with exactly solvable topological hinge and corner states, *Phys. Rev. B* **97**, 241405(R) (2017).

32. Ni, X., Weiner, M., Alù, A. & Khanikaev, A. B. Observation of higher-order topological acoustic states protected by generalized chiral symmetry, *Nat. Materials* **18**, 113-120 (2019).

33. Xue, H., Yang, Y., Gao, F., Chong, Y. & Zhang, B. Acoustic higher-order topological insulator on a kagome lattice, *Nat. Materials* **18**, 108-112 (2019).

34. Noh, J., Benalcazar, W. A., Huang, S., Collins, M. J., Chen, K. P., Hughes, T. L. & Rechtsman, M. C. Topological protection of photonic mid-gap defect modes, *Nat. Photonics* **12**, 408 (2018).

35. Xie, B.-Y., Wang, H.-F., Wang, H.-X., Zhu, X.-Y., Jiang, J.-H., Lu, M.-H. & Chen, Y.-F. Second-order photonic topological insulator with corner states, *Phys. Rev. B* **98**, 205147 (2018).

36. Peterson, C. W., Benalcazar, W. A., Hughes, T. L. & Bahl, G. A quantized microwave quadrupole insulator with topologically protected corner states, *Nature* **555**, 346 (2018).

37. Serra-Garcia, M., Peri, V., Süsstrunk, R., Bilal, O. R., Larsen, T., Villanueva, L. G. & Huber, S. D. Observation of a phononic quadrupole topological insulator, *Nature* **555**, 342 (2018).

38. Mittal, S., Orre, V. V., Zhu, G., Gorlach, M. A., Poddubny, A. & Hafezi, M. Photonic quadrupole topological phases, arXiv 1812.09304 (2018).

39. Imhof, S. Berger, C., Bayer, F., Brehm, J., Molenkamp, L. W., Kiessling, T., Schindler, F., Lee, C. H., Greiter, M., Neupert, T. & Thomale, R. Topolectrical-circuit realization of topological corner modes, *Nat. Phys.* **14**, 925 (2018).

40. Yuan, L., Lin, Q., Xiao, M. & Fan, S. Synthetic dimension in photonics, *Optica* **5**, 1396 (2018).

41. Yuan, L., Shi, Y. & Fan, S. Photonic gauge potential in a system with a synthetic frequency dimension, *Opt. Lett.* **41**, 741 (2016).

42. Ozawa, T., Price, H. M., Goldman, N., Zilberberg, O. & Carusotto, I. Synthetic dimensions in integrated photonics: From optical isolation to four-dimensional quantum Hall physics, *Phys. Rev. A* **93**, 043827 (2016).

43. Yuan, L., Xiao, M., Lin, Q., & Fan, S. Synthetic space with arbitrary dimensions in a few rings undergoing dynamic modulation, *Phys. Rev. B* **97**, 104105 (2018).

44. Luo, X.-W., Zhou, X., Xu, J.-S., Li, C.-F., Guo, G.-C., Zhang, C. & Zhou, Z.-W. Synthetic-lattice enabled all-optical devices based on orbital angular momentum of light, *Nat. Commun.* **8**, 16097 (2017).

45. Zhou, X.-F., Luo, X.-W., Wang, S., Guo, G.-C., Zhou, X., Pu, H., & Zhou, Z.-W. Dynamically Manipulating Topological Physics and Edge Modes in a Single

Degenerate Optical Cavity, *Phys. Rev. Lett.* **118**, 083603 (2017).

46. Kraus, Y. E., Lahini, Y., Ringel, Z., Verbin, M. & Zilberberg, O. Topological states and adiabatic pumping in quasicrystals, *Phys. Rev. Lett.* **109**, 106402 (2012).

47. Verbin, M., Zilberberg, O., Kraus, Y. E., Lahini, Y. & Silberberg, Y. Observation of topological phase transitions in photonic quasicrystals, *Phys. Rev. Lett.* **110**, 076403 (2013).

48. Zilberberg, O., Huang, S., Guglielmon, J., Wang, M., Chen, K. P., Kraus, Y. E. & Rechtsman, M. C. Photonic topological boundary pumping as a probe of 4D quantum Hall physics, *Nature* **553**, 59–62 (2018).

49. Wang, Q., Xiao, M., Liu, H., Zhu, S. & Chan, C. T. Optical interface states protected by synthetic Weyl points, *Phys. Rev. X* **7**, 031032 (2017).

50. Yuan, L. & Fan, S. Bloch oscillation and unidirectional translation of frequency in a dynamically modulated ring resonator, *Optica* **3**, 1014–1018 (2016).

51. Yuan, L., Lin, Q., Zhang, A., Meng, X., Chen, X. & Fan, S. Photonic Gauge Potential in One Cavity with Synthetic Frequency and Orbital Angular Momentum Dimensions, *Phys. Rev. Lett.* **122**, 083903 (2019).

52. Lin, Q., Xiao, M., Yuan, L. & Fan, S. Photonic Weyl point in a two-dimensional resonator lattice with a synthetic frequency dimension, *Nat. Commun.* **7**, 13731 (2016).

53. Lin, Q., Sun, X.-Q., Xiao, M., Zhang, S.-C. & Fan, S. Constructing three-dimensional photonic topological insulator using two-dimensional ring resonator lattice with a synthetic frequency dimension, arXiv: 1802.02597 (2018).

54. Lustig, E., Weimann, S., Plotnik, Y., Lumer, Y., Bandres, M. A., Szameit, A. & Segev, M. Photonic topological insulator in synthetic dimensions, *Nature* **567**, 356-360 (2019).

55. Blanco-Redondo, A., Bell, B., Oren, D., J. Eggleton, B. & Segev, M. Topological protection of biphoton states, *Science* **362**, 568-571 (2018).

**Acknowledgements**

This work was supported by the National key R&D Program of China under Grant No. 2017YFA0303800 and the National Natural Science Foundation of China through

Grants No. 91850205 and No.61421001.

## Author contributions

All authors contributed to discussions and project development. W.X. Z. provided the theoretical design and analysis. X. D. Z. initiated and designed this research project.

## Competing interests

The authors declare no competing financial interests.